\documentclass{article}
\usepackage[]{graphicx}
\usepackage{epsfig}
\usepackage{amsmath, amstext, amsfonts}
\usepackage{amssymb}
\usepackage{multicol}
\usepackage[latin1]{inputenc}

\begin{document}

\title{Bloch Solutions of Periodic Dirac Equations in SPPS Form\footnote{Operator Theory: Advances and Applications {\bf 220}, 153-162 (2012). Springer, Basel AG.}}
\author{K.V. Khmelnytskaya$^1$ and H.C. Rosu$^2$\\
%EndAName
$^1$ Applied Mathematics, Faculty of Engineering,\\ Autonomous University of Queretaro,
Centro Universitario,\\ Cerro de las Campanas s/n C.P. 76010 Santiago de Queretaro, Qro., Mexico\\
$^2$ IPICYT, Instituto Potosino de Investigacion Cientifica y Tecnologica,\\Apdo Postal 3-74 Tangamanga, 78231 San Luis Potos\'{\i}, S.L.P., Mexico}
\date{}
\maketitle

\noindent Subjclass: Primary 34B24; Secondary 34C25 


\noindent File: Iwota-09-Kira-HC.tex

\begin{abstract}
\noindent We provide the representation of quasi-periodic solutions of periodic Dirac
equations in terms of the spectral parameter power series (SPPS) recently
introduced by V.V. Kravchenko \cite{Krav2008,Krav2009,KP}. We also give the
SPPS form of the Dirac Hill discriminant
under the Darboux nodeless transformation using the SPPS form of the discriminant
and apply the results to one of Razavy's quasi-exactly solvable periodic potentials.
\end{abstract}

%%% ----------------------------------------------------------------------
\maketitle

\section{Introduction}

The connections between the Dirac equation and the Schr\"odinger equation are
known since a long time ago \cite{biedenharn} and have been strengthen in the
supersymmetric context soon after the advent of supersymmetric quantum
mechanics in 1981 \cite{sukumar,hughes,cooper,nogami}. There are currently
interesting applications of this approach in condensed matter physics
\cite{JP07,kr,kuru}. In this work, we are interested in the same connection in
the case of periodic potentials, see e.g. \cite{samsonov}. We here write the
Dirac Bloch solutions in Kravchenko form (power series in the spectral
parameter) and also the Dirac Hill discriminant in the same form and apply the
results to an interesting quasi-exactly solvable periodic potential.

\section{Schr\"odinger equations of Hill type}

The Schr\"odinger differential equation
%...........%
\begin{equation}
L\left[ f(x,\lambda)\right]  =-f^{\prime\prime}(x,\lambda)+q(x)f(x,\lambda
)=\lambda f(x,\lambda)\label{eq}%
\end{equation}
%.............
with $T$-periodic real-valued potential $q(x)$ assumed herewith a continuous
bounded function and $\lambda$ a real parameter is known as of Hill type. We
begin by recalling some necessary definitions and basic properties associated
with the equation (\ref{eq}) from the Floquet (Bloch) theory. For more details
see, e.g., \cite{Magnus,Eastham}.
%In what follows we assume that $q(x)$ is a continuous bounded functions.

For each $\lambda$ there exists a fundamental system of solutions, i.e., two
linearly independent solutions of (\ref{eq}), $f_{1}(x,\lambda)$ and
$f_{2}(x,\lambda)$, which satisfy the initial conditions%
\begin{equation}
f_{1}(0,\lambda)=1,\quad f_{1}^{\prime}(0,\lambda)=0,\quad f_{2}%
(0,\lambda)=0,\quad f_{2}^{\prime}(0,\lambda)=1.\label{cond}%
\end{equation}
Then the Hill discriminant associated with equation (\ref{eq}) is defined as a
function of $\lambda$ as follows
%............%
\[
D(\lambda)=f_{1}(T,\lambda)+f_{2}^{\prime}(T,\lambda).
\]
%............
The importance of $D(\lambda)$ stems from the easiness of describing the
spectrum of the corresponding equation by its means, namely \cite{Magnus}:

\medskip

\begin{quote}
(1) sets $\{\lambda_{i}\}$ for which $\left\vert D(\lambda)\right\vert \leq2$
form the allowed bands or stability intervals,\newline(2) sets $\{\lambda
_{j}\}$ for which $\left\vert D(\lambda)\right\vert > 2$ form the forbidden
bands or instability intervals,\newline(3) sets $\{\lambda_{k}\}$ for which
$\left\vert D(\lambda)\right\vert =2$ form the band edges and represent the
discrete part of the spectrum.
\end{quote}

\medskip

\noindent Furthermore, when $D(\lambda)=2$ equation (\ref{eq}) has a periodic
solution with the period $T$ and when $D(\lambda)=-2$ it has an aperiodic
solution, i.e. $f(x+T)=-f(x)$. The eigenvalues $\lambda_{n}$, $n=0,1,2,...$
form an infinite sequence $\lambda_{0}<\lambda_{1}\leq\lambda_{2}<\lambda
_{3}\leq\lambda_{4}...$, and an important property of the minimal eigenvalue
$\lambda_{0}$ is the existence of a corresponding periodic nodeless solution
$u(x,\lambda_{0})$ \cite{Magnus}. The solutions of (\ref{eq}) are not periodic
in general, and one of the important tasks is the construction of
quasiperiodic solutions defined by $f_{\pm}(x+T)=\beta_{\pm}(\lambda)f_{\pm
}(x)$. Here, we use James' matching procedure \cite{James} that employs the
fundamental system of solutions, $f_{1}(x,\lambda)$ and $f_{2}(x,\lambda)$, in
the construction of the quasiperiodic solutions as follows
%..............%
\begin{equation}
\label{bloch}f_{\pm}(x,\lambda)=\beta_{\pm}^{n}(\lambda)\left[ f_{1}%
(x-nT,\lambda)+\alpha_{\pm}f_{2}(x-nT,\lambda)\right]  ,\quad\left\{
\begin{array}
[c]{c}%
nT\leq x<(n+1)T\\
n=0,\pm1,\pm2,...~
\end{array}
\right.  ,
\end{equation}
%...............
where $\alpha_{\pm}$ are given by \cite{James}
%the roots of the simple algebraic equation $f_{2}(T,\lambda)\alpha^{2}+(f_{1}(T,\lambda
%)-f_{2}^{\prime}(T,\lambda))\alpha-f_{1}^{\prime}(T,\lambda)=0$.%
\begin{equation}
\alpha_{\pm}=\frac{f_{2}^{\prime}(T,\lambda)-f_{1}(T,\lambda)\mp\left(
D^{2}(\lambda)-4\right)  ^{\frac{1}{2}}}{2f_{2}(T,\lambda)} ~.\label{alfa}%
\end{equation}
The Bloch factors $\beta_{\pm}(\lambda)$ are a measure of the rate of increase
(or decrease) in magnitude of the linear combination of the fundamental system
when one goes from the left end of the cell to the right end, i.e.,
%............%
\[
\beta_{\pm} (\lambda)=\frac{f_{1}(T,\lambda)+\alpha_{\pm}f_{2}(T,\lambda
)}{f_{1}(0,\lambda)+\alpha_{\pm}f_{2}(0,\lambda)}~.
\]
The values of $\beta_{\pm}(\lambda)$ are directly related to the Hill
discriminant$,\beta_{\pm}(\lambda)=\frac{1}{2}(D(\lambda)\mp\sqrt
{D^{2}(\lambda)-4})$, and obviously at the band edges $\beta_{+}=\beta_{-}%
=\pm1$ for $D(\lambda)=\pm2$, respectively.

\section{SPPS representation for solutions of the one-dimensional Dirac
equation}

We consider the following Dirac equation%
\begin{equation}
L[W]=\left[  -i \sigma_{y}d_{x}+\sigma_{x}\Phi(x)\right]  W=\omega
W,\label{Dirac}%
\end{equation}
where the scalar potential $\Phi(x)$ is periodic function with period $T$, $W
$ is the spinor $W=\left(
\begin{array}
[c]{c}%
f\\
g
\end{array}
\right)  $ and $\sigma_{x}$, $\sigma_{y}$ are the Pauli matrices $\sigma
_{x}=\left(
\begin{array}
[c]{cc}%
0 & 1\\
1 & 0
\end{array}
\right)  $ and $\sigma_{y}=\left(
\begin{array}
[c]{cc}%
0 & -i\\
i & 0
\end{array}
\right) .$

The uncoupled Schr\"{o}dinger equations derived from equation (\ref{Dirac})
are%
\begin{align}
(-d_{x}+\Phi)(d_{x}+\Phi)f  &  =\lambda f,\label{a}\\
(d_{x}+\Phi)(-d_{x}+\Phi)g  &  =\lambda g,\label{b}%
\end{align}
where $\lambda=\omega^{2}$ is the spectral parameter. It is clear that the
solutions $f$ and $g$ are related by the following relationship
\begin{equation}
(d_{x}+\Phi)f=\omega g,\label{c}%
\end{equation}
therefore with the solution $f$ at hand, we can construct the solution $g$ immediately.

We start with equation (\ref{a}). Notice that the solution $u$ of the equation
(\ref{a}) for $\lambda=0$ can be obtained as follows $u(x)=e^{-\int\Phi(x)dx}$
\ and $u(x)$ is a nodeless periodic function with the period $T$ if
$\Phi(x)\in\mathbf{C}^{1}$ and $\int_{0}^{T}\Phi(x)dx=0$.

Once having the function $u(x)$ the solutions $f_{1}(x,\lambda)$ and
$f_{2}(x,\lambda)$ of (\ref{a}), (\ref{cond}) for all values of the parameter
$\lambda$ can be given using the SPPS method \cite{Krav2008} .
\begin{align}
f_{1}(x,\lambda)  &  =\frac{u(x)}{u(0)}\widetilde{\Sigma}_{0}(x,\lambda
)+u^{\prime}(0)u(x)\Sigma_{1}(x,\lambda),\nonumber\\
& \label{f1 f2}\\
f_{2}(x,\lambda)  &  =-u(0)u(x)\Sigma_{1}(x,\lambda).\nonumber
\end{align}
%............
The functions $\widetilde{\Sigma}_{0}$ and $\Sigma_{1}$ are the spectral
parameter power series
%........%
\[
\widetilde{\Sigma}_{0}(x,\lambda)=\sum_{\,n=0}^{\infty}\widetilde{X}%
^{(2n)}(x)\lambda^{n},\quad\Sigma_{1}(x,\lambda)=\sum_{n=1}^{\infty}%
X^{(2n-1)}(x)\lambda^{n-1}~,
\]
%.........
where the coefficients $\widetilde{X}^{(n)}(x)$, $X^{(n)}(x)$ are given by the
following recursive relations
%...............%
\[
\widetilde{X}^{(0)}\equiv1,\qquad X^{(0)}\equiv1,
\]
%...........
%

\begin{equation}
\tilde{X}^{(n)}(x)=%
\begin{cases}
\int_{0}^{x}\tilde{X}^{(n-1)}(\xi)u^{2}(\xi)d\xi\qquad\mathrm{for}%
\,\mathrm{an}\,\mathrm{odd}\,n\\
\\
-\int_{0}^{x}\tilde{X}^{(n-1)}(\xi)\frac{d\xi}{u^{2}(\xi)}\qquad
\ \ \ \ \mathrm{for}\,\mathrm{an}\,\mathrm{even}\,n
\end{cases}
\label{K1}%
\end{equation}

\bigskip%

\begin{equation}
X^{(n)}(x)=%
\begin{cases}
-\int_{0}^{x}X^{(n-1)}(\xi)\frac{d\xi}{u^{2}(\xi)}\qquad\ \ \ \ \ \mathrm{for}%
\,\mathrm{an}\,\mathrm{odd}\,n\\
\\
\int_{0}^{x}X^{(n-1)}(\xi)u^{2}(\xi)d\xi\qquad\ \mathrm{for}\,\mathrm{an}%
\,\mathrm{even}\,n~.
\end{cases}
\label{K2}%
\end{equation}

One can check by a straightforward calculation that the solutions $f_{1}$ and
$f_{2}$ fulfill the initial conditions (\ref{cond}), for this the following
relations are useful
%..........9%
\begin{equation}
\left(  \widetilde{\Sigma}_{0}(x,\lambda)\right)  _{x}^{\prime}=-\frac
{\widetilde{\Sigma}_{1}(x,\lambda)}{u^{2}(x)},\ \text{where}\,\,\widetilde
{\Sigma}_{1}(x,\lambda)=\sum_{\,n=1}^{\infty}\widetilde{X}^{(2n-1)}%
(x)\lambda^{n}\label{S1}%
\end{equation}
%.............10
and%
\begin{equation}
\left(  \Sigma_{1}(x,\lambda)\right)  _{x}^{\prime}=-\frac{\Sigma
_{0}(x,\lambda)}{u^{2}(x)},\ \text{where}\,\,\Sigma_{0}(x,\lambda)=\sum
_{n=0}^{\infty}X^{(2n)}(x)\lambda^{n}.\label{S2}%
\end{equation}
The pair of linearly independent solutions $g_{1}(x,\lambda)$ and
$g_{2}(x,\lambda)$\ of (\ref{b}) can be obtained directly from the solutions
(\ref{f1 f2}) by means of (\ref{c}). We additionally take the linear
combinations in order that the solutions $g_{1}(x,\lambda)$ and $g_{2}%
(x,\lambda)$\ satisfy the initial conditions $g_{1}(0,\lambda)=g_{2}^{\prime
}(0,\lambda)=1$ and $g_{1}^{\prime}(0,\lambda)=g_{2}(0,\lambda)=0$%

\begin{align}
g_{1}(x,\lambda)  &  =\frac{u(0)}{u(x)}\Sigma_{0}(x,\lambda)-\frac{\Phi
(0)}{\lambda u(0)u(x)}\widetilde{\Sigma}_{1}(x,\lambda),\nonumber\\
& \label{g1 g2}\\
g_{2}(x,\lambda)  &  =\frac{1}{\lambda u(0)u(x)}\widetilde{\Sigma}%
_{1}(x,\lambda).\nonumber
\end{align}

Thus, the two spinor solutions of the Dirac equation (\ref{Dirac}) are given
by%
\[
W_{1}=\left(
\begin{array}
[c]{c}%
f_{1}\\
g_{1}%
\end{array}
\right)  ,\text{ and }W_{2}=\left(
\begin{array}
[c]{c}%
f_{2}\\
g_{2}%
\end{array}
\right)
\]
and these solutions satisfy the following initial conditions%
\[
W_{1}(0)=\left(
\begin{array}
[c]{c}%
1\\
0
\end{array}
\right)  ,\text{ and }W_{2}(0)=\left(
\begin{array}
[c]{c}%
0\\
1
\end{array}
\right)  .
\]

\subsection{\bigskip Bloch solutions and Hill's discriminant}

The second order differential equations (\ref{a}) and (\ref{b}) have periodic
potentials $V_{1,2}=\Phi^{2}\mp\Phi^{\prime}$, correspondingly. The important
tasks for this case are the construction of the Bloch solutions which are
subject to the Bloch condition $f(x+T)=e^{ikT}f(x)$ ($k$, a wave number) and the description of the spectrum.

In \cite{KR} the SPPS representations of Hill discriminants $D_{f}(\lambda)$
and $D_{g}(\lambda)$ associated with the equations (\ref{a}) and (\ref{b})
were obtained in the form
%Thus we have, respectively, $D_{f}(\lambda)$ and $D_{g}(\lambda)$%
\begin{align*}
D_{f}(\lambda)  &  =\frac{u(T)}{u(0)}\widetilde{\Sigma}_{0}(T,\lambda
)+\frac{u(0)}{u(T)}\Sigma_{0}(T,\lambda)\\
&  +\left(  u^{\prime}(0)u(T)-u(0)u^{\prime}(T)\right)  \Sigma_{1}%
(T,\lambda)~,
\end{align*}%
\begin{align*}
D_{g}(\lambda)  &  =\frac{u(0)}{u(T)}\Sigma_{0}(T,\lambda)+\frac{u(T)}%
{u(0)}\widetilde{\Sigma}_{0}(T,\lambda)+\frac{1}{\left(  \Delta\lambda\right)
u^{2}(0)u^{2}(T)}\left(  u^{\prime}(0)u(T)-\right. \\
&  \left.  -u^{\prime}(T)u(0)\right)  \widetilde{\Sigma}_{1}(T,\lambda).
\end{align*}
It is clear that since $u(x)$ is a $T$-periodic function ($u(0)=u(T)$ ) the
expression in brackets in the above formulae vanishes. Now\ writing the
explicit expressions for $\widetilde{\Sigma}_{0}(T,\lambda,\lambda_{0})$ and
$\Sigma_{0}(T,\lambda,\lambda_{0})$, a representation for Hill's discriminant
associated with (\ref{a}) and (\ref{b}) is the following%
\begin{equation}
D_{f}(\lambda)=D_{g}(\lambda)=\sum_{n=0}^{\infty}\left(  \tilde{X}%
^{(2n)}(T)+X^{(2n)}(T)\right)  \lambda^{n}\text{.}\label{D}%
\end{equation}
Equations (\ref{a}) and (\ref{b}) are isospectral and we obtain the Hill
discriminant associated with the Dirac equation (\ref{Dirac}). We formulate
this result as the following theorem:\\

\noindent {\bf Theorem}.
{\em Let $\Phi(x)\in\mathbf{C}^{1}$ be a $T$-periodic function which satisfies the
condition $\int_{0}^{T}\Phi(x)dx=0$. Then the Hill discriminant for
(\ref{Dirac}) has the form
\[
D_{W}(\omega)=\sum_{n=0}^{\infty}\left(  \tilde{X}^{(2n)}(T)+X^{(2n)}%
(T)\right)  \omega^{2n}~,
\]
where $\tilde{X}^{(2n)}$ and $X^{(2n)}$ are calculated according to (\ref{K1})
and (\ref{K2}), $u=e^{-\int\Phi(x)dx}$ and the series converges uniformly on
any compact set of values of $\omega$.}\\

In order to construct the Bloch solutions for the Dirac equation (\ref{Dirac})
we use the solutions (\ref{f1 f2}) and (\ref{g1 g2}) and apply the procedure
of James \cite{James}. Notice that because the Hill discriminants for the
equations (\ref{a}) and (\ref{b}) are identical the Bloch factors for both
equations are equal. The so-called self-matching solutions for the equations
(\ref{a}) and (\ref{b}), are correspondingly%
\[
F_{\pm}(x,\lambda)=f_{1}(x,\lambda)+a_{\pm}f_{2}(x,\lambda)\text{ and }G_{\pm
}(x,\lambda)=g_{1}(x,\lambda)+b_{\pm}g_{2}(x,\lambda),
\]
where $a_{\pm}$ and $b_{\pm}$ are calculated by the formula (\ref{alfa}) with
the corresponding fundamental system of solutions (\ref{f1 f2}) and
(\ref{g1 g2}). By means of $F_{\pm}$ and $G_{\pm}$ we write the self-matching
spinor solution of the equation (\ref{Dirac})%
\[
w_{\pm}(x,\lambda)=\left(
\begin{array}
[c]{c}%
F_{\pm}(x,\lambda)\\
G_{\pm}(x,\lambda)
\end{array}
\right)  .
\]
\ Finally, the Bloch solutions of the equation (\ref{Dirac}) take the form%
\[
W_{\pm}(x,\lambda)=\beta_{\pm}^{n}(\lambda)\left(  w_{\pm}(x-nT,\lambda
)\right)  ,\quad\left\{
\begin{array}
[c]{c}%
nT\leq x<(n+1)T\\
n=0,\pm1,\pm2,...~.
\end{array}
\right.
\]

\section{Numerical calculation of eigenvalues based on the SPPS form of Hill's
discriminant}

As is well known \cite{Magnus}, the zeros of the functions $D(\lambda)\mp2$
represent eigenvalues of the corresponding Hill operator with periodic and
aperiodic boundary conditions, respectively. In this section, we show that
besides other possible applications the representation (\ref{D}) gives us an
efficient tool for the calculation of the discrete spectrum of a periodic
Dirac operator.

The first step of the numerical realization of the method consists in
calculation of the functions $\tilde{X}^{(n)}$ and $X^{(n)}$ given by
(\ref{K1}) and (\ref{K2}), respectively. This construction is based on the
eigenfunction $u(x)$. Next, by truncating the infinite series for $D(\lambda
)$\ (\ref{D}) we obtain a polynomial in $\lambda$
%.........%
\begin{align}
D_{N}(\lambda)  &  =\sum_{n=0}^{N}\left(  \tilde{X}^{(2n)}(T)+X^{(2n)}%
(T)\right)  \lambda^{n}\label{DN}\\
&  =2+\sum_{n=1}^{N}\left(  \tilde{X}^{(2n)}(T)+X^{(2n)}(T)\right)
\lambda^{n}.\nonumber
\end{align}
%.........
The roots of the polynomials $D_{N}(\lambda)\mp2$ give us the eigenvalues
corresponding to equation (\ref{eq}) with periodic and aperiodic boundary
conditions, respectively.

\bigskip

As an example we consider the Dirac equation (\ref{Dirac}) with the scalar
potential
\[
\Phi(x)=\sin2x\left[  \frac{\xi}{2}-\frac{2A(\xi)}{\xi-A(\xi) \cos2x}\right]
~,
%\quad A(\xi)=\left(1-\sqrt{1+\xi^{2}}\right)
\]
with $A(\xi)=\left( 1-\sqrt{1+\xi^{2}}\right) $ and $\xi$ a real positive
parameter. This scalar potential satisfies the conditions of theorem 3.1. The
corresponding second order differential equations are%
\begin{align*}
-d_{x}^{2}f+V_{1}f  &  =\lambda f,\\
-d_{x}^{2}g+V_{2}g  &  =\lambda g,
\end{align*}
where the Schr\"odinger potential
\begin{equation}\label{V1}
V_{1}(x) =\frac{\xi^{2}}{8}\left(  1-\cos4x\right)  -3\xi\cos2x~,
\end{equation}
is the case $m=2$ in the quasi-exactly solvable family of the so-called trigonometric
Razavy potentials \cite{raz}, $V_{R} =\frac{\xi^{2}}{8}\left(  1-\cos4x\right)  -(m+1)\xi\cos2x$.
For a given integer $m$, if $\xi< 2(m+1)$ the potentials $V_R(x)$ are of single-well periodic type
and if $\xi>2(m+1)$ they are of double-well periodic type.
\begin{equation}\label{V2}
V_{2}(x) =V_{1}(x) +4\cos2x\left( \frac{\xi}{2}-\frac{2A(\xi)}{\xi-A(\xi)
\cos2x}\right)  +\frac{8A(\xi) \sin^{2}2x}{\left(  \xi-A(\xi) \cos2x\right)
^{2}}%
\end{equation}
%\end{align*}
is the supersymmetric partner potential and therefore it is also quasi-exactly
solvable. The Schr\"{o}dinger equations with these potentials can be used for
the description of torsional oscillations of certain molecules \cite{raz}.
Plots of the potentials $V_1(x)$ and $V_2(x)$ are displayed in Fig.~1 for two values of $\xi$.

%..... Figure 1
\begin{figure}[h]
%\caption{\textsl{The Razavy potentials $V_{1}$ (solid lines) given by
%(\ref{V1}) and its partner potentials $V_{2}$ (dashed lines) as given by
%(\ref{V2}). The thinner curves correspond to $\xi=1$ and the thick curves are
%for $\xi=2$. }}%
%\label{razF1}
%%\centering
%\par
\begin{center}
\includegraphics[width= 14.5 cm, height=6.5 cm]{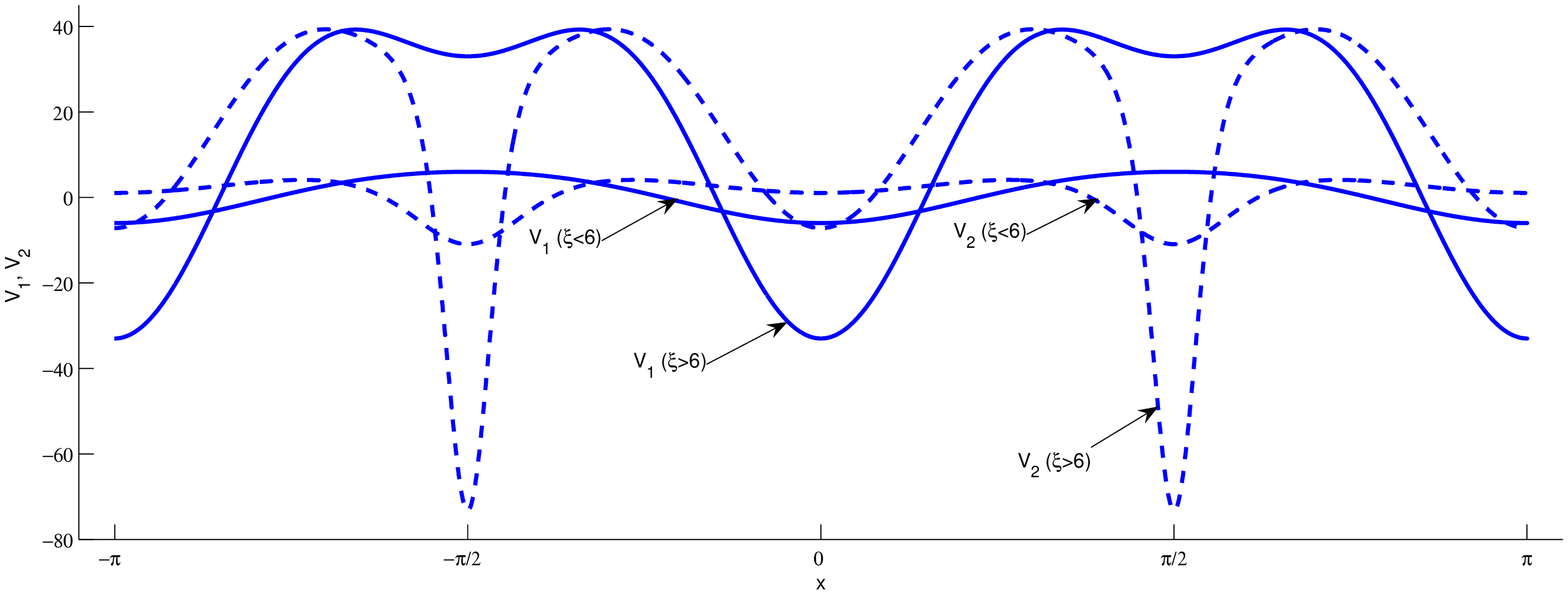}
\caption{\textsl{The Razavy potentials $V_{1}$ (solid lines) for $\xi=2$ and $\xi=11$ given by
(\ref{V1}) and its partner potentials $V_{2}$ (dashed lines) as given by
(\ref{V2}) for the same values of $\xi$. }}%
\label{razF1}
%\centering
\end{center}
\end{figure}

The computer algorithm was implemented in Matlab 2006. The recursive
integration required for the construction of $\tilde{X}_{0}^{(n)}$,
$X_{0}^{(n)}$, $\tilde{X}^{(n)}$ and $X^{(n)}$ was done by representing the
integrand through a cubic spline using the $\emph{spapi}$ routine with a
division of the interval $[0,\pi]$ into $5000$ subintervals and integrating
using the $\emph{fnint}$ routine. Next, the zeros of $D_{N}(\lambda)\pm
2$\ were calculated by means of the $\emph{fnzeros}$ routine.

%\bigskip

In the following tables, the eigenvalues were calculated employing the SPPS representation (\ref{D}) for four different values of the parameter
$\xi$. The first two values are below the threshold value $\xi _{\rm thr}=6$ for $m=2$ from single-well to double-well types of Razavy's potentials while the last two values are above this threshold value. For comparison, we use the eigenvalues given analytically by Razavy in terms of the parameter $\xi$
as follows \cite{raz}%
\[
\lambda_{0}=2\left(  1-\sqrt{1+\xi^{2}}\right)  ,\quad\lambda_{3}%
=4,\quad\lambda_{4}=2\left(  1+\sqrt{1+\xi^{2}}\right) ~.
\]

$
%\begin{tabular}
%[b]{|c|}\hline
%$\xi=1$  \\\hline
%\end{tabular}%
\begin{tabular}
[b]{|l|l|l|}\hline
& \qquad\quad$\xi=1$ & \qquad\quad$\xi=1$\\\hline
$n$ & \qquad$\lambda_{n}\ \text{(SPPS\thinspace)}$ & \qquad$\lambda
_{n}\ \text{(Ref. \cite{raz}\thinspace)}$\\\hline
$0$ & $-0.828427124746190$ & $-0.828427124746190$\\\hline
$1$ & $-0.628906956748252$ & \\\hline
$2$ & $2.315132548422588$ & \\\hline
$3$ & $3.999991462865745$ & $4$\\\hline
$4$ & $4.828420096225068$ & $4.828427124746190$\\\hline
$5$ & $9.238264469324272$ & \\\hline
$6$ & $9.294265517212145$ & \\\hline
\end{tabular}
\ \ \qquad$

$%
\begin{tabular}
[b]{|l|l|l|}\hline
& \qquad\quad$\xi=2$ & \qquad\quad$\xi=2$\\\hline
$n$ & \qquad$\lambda_{n}\ \text{(SPPS\thinspace)}$ & \qquad$\lambda
_{n}\ \text{(Ref. \cite{raz})}\,$\\\hline
$0$ & $-2.472135954999580$ & $-2.472135954999580$\\\hline
$1$ & $-2.428136886851045$ & \\\hline
$2$ & $3.184130151531468$ & \\\hline
$3$ & $4.000004180961838$ & $4$\\\hline
$4$ & $6.472138385406806$ & $6.472135954999580$\\\hline
$5$ & $9.864117523158974$ & \\\hline
$6$ & $10.253256926576858$ & \\\hline
\end{tabular}
\ \ $

%\bigskip

%\bigskip

$%
\begin{tabular}
[b]{|l|l|l|}\hline
& \qquad\quad$\xi=11$ & \qquad\quad$\xi=11$\\\hline
$n$ & \qquad$\lambda_{n}\ \text{(SPPS\thinspace)}$ & \qquad$\lambda
_{n}\ \text{(Ref. \cite{raz})}\,$\\\hline
$0$ & $-20.090722034374522$ & $-20.090722034374522$\\\hline
$1$ & $-20.090721031408926$ & \\\hline
$2$ & $3.999728397824670$ & \\\hline
$3$ & $4.000000543012631$ & $4$\\\hline
$4$ & $24.092379855485746$ & $24.090722034374522$\\\hline
$5$ & $24.125593160436161$ & \\\hline
$6$ & $36.212102534969766$ & \\\hline
\end{tabular}
\ \ \qquad$

$%
\begin{tabular}
[b]{|l|l|l|}\hline
& \qquad\quad$\xi=20$ & \qquad\quad$\xi=20$\\\hline
$n$ & \qquad$\lambda_{n}\ \text{(SPPS\thinspace)}$ & \qquad$\lambda
_{n}\ \text{(Ref. \cite{raz})}\,$\\\hline
$0$ & $-38.049968789001575$ & $-38.049968789001575$\\\hline
$1$ & $-38.049968788934475$ & \\\hline
$2$ & $3.999999942823312$ & \\\hline
$3$ & $3.999999999630503$ & $4$\\\hline
$4$ & $42.050313148383374$ & $42.049968789001575$\\\hline
$5$ & $42.050347742353317$ & \\\hline
$6$ & $74.691604620863302$ & \\\hline
\end{tabular}
$

\noindent In Fig. 2, we display the plots of the Hill discriminants for
the values of the Razavy parameter $\xi=1$, $\xi=2$, and $\xi=3$, respectively. In general, these plots contain damped oscillations with higher amplitudes at higher $\xi$. On the other hand, getting the spectrum in $\lambda$ is equivalent with having the eigenvalues $\omega_{n}=\pm \sqrt{\lambda_n}$ of the Dirac system under consideration.
%..... Figure 2
\begin{figure}[h]
%\centering
\par
\begin{center}
\includegraphics[width= 14.5 cm, height=6.5 cm]{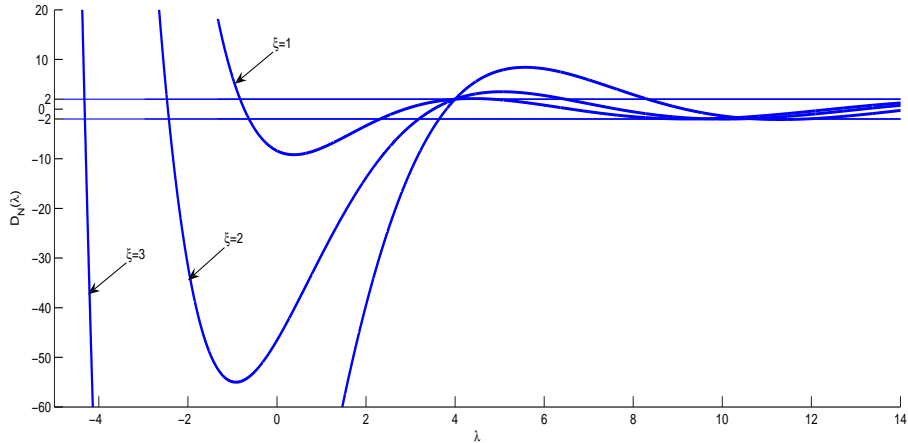}
\end{center}
\caption{\textsl{The polynomial $D_{N}(\lambda)$ for the Hill equations with Razavy's partner potentials for three values
of the parameter $\xi$ calculated by means of formula (\ref{DN}) for $N=100$. The first minimum of the discriminant in this plot, i.e., for $\xi=3$, goes down to -260.9 at $\lambda =-2.469$.}}%
\label{razF2}%
\end{figure}

\section{Conclusions}

In summary, in this work we presented the SPPS form of the quasi-periodic
(Bloch) solutions of periodic one-dimensional Dirac operators as well as of
the Hill discriminant. We applied the obtained results to the Dirac system
with the periodic scalar potential that leads to one of Razavy's quasi-exactly
solvable periodic potentials.

%------------------------------------------------------------------------

\subsection*{Acknowledgment}

The first author thanks CONACyT for a postdoctoral fellowship allowing her to
work in IPICyT.
%------------------------------------------------------------------------

\end{document}